# Multiple ionization of argon via xuv-photon absorption induced by 20-gigawatt high-harmonic pulses

A. Nayak[1*], I. Orfanos[1,2*], I. Makos[1,2*], M. Dumergue[3*], S. Kühn[3*], E. Skantzakis[1], B. Bodi[4], K. Varju[3,5], C. Kalpouzos[1], H. I. B. Banks[6], A. Emmanouilidou[6], D. Charalambidis[1,2,3†] and P. Tzallas[1†]

[1]*Foundation for Research and Technology - Hellas, Institute of Electronic Structure & Laser, PO Box 1527, GR71110 Heraklion (Crete), Greece*
[2]*Department of Physics, University of Crete, PO Box 2208, GR71003 Heraklion (Crete), Greece*
[3]*ELI-ALPS, ELI-Hu Kft., Dugonics ter 13, H-6720 Szeged Hungary*
[4]*MTA "Lendület" Ultrafast Nanooptics Group, Wigner Research Center for Physics, 1121 Budapest, Hungary*
[5]*Department of Optics and Quantum Electronics, University of Szeged, Szeged Hungary*
[6]*Department of Physics and Astronomy, University College London, Gower Street, London WC1E 6BT, United Kingdom*

[*] Equally contributed authors
[†] Corresponding authors: chara@iesl.forth.gr and ptzallas@iesl.forth.gr

We report the observation of multiple ionization of Argon through multi-XUV-photon absorption induced by an unprecedentedly powerful laser driven high-harmonic-generation source. Comparing the measured intensity dependence of the yield of the different Argon charge states with numerical calculations we can infer the different channels -direct and sequential- underlying the interaction. While such studies were feasible so far only with FEL sources, this work connects highly-non-linear-XUV-processes with the ultra-short time scales, inherent to the harmonic pulses, and highlights the advanced perspectives of emerging large scale laser research infrastructures.

Multiphoton processes trace back to 1931 [1]. One third of a century later, the invention of lasers has allowed their observation [2], followed by several decades of flourishing multi-photon and strong field science. Another third of century later two-photon processes made their debut in the XUV spectral region [3], paving the way to XUV-pump-XUV-probe studies in the temporal scale of 1 femtosecond (*fs*) [4] and below. Instrumental to these studies is high-order harmonic generation (HHG), which apart from its importance in understanding strong-field laser-atom interactions [5-7], led to the development of coherent XUV sources operating down to the ~ 10 nm spectral range with pulse durations in the attosecond (*asec*) regime [8, 9, 10-14]. While HHG/attosecond sources induced so far few photon (mainly two-photon) processes (mainly single- or double-ionization), FEL sources have achieved the production of high charge states of ions through absorption of many XUV [15] or even x-ray [16-18] photons. Attaining insight into such processes is considered of central importance [15-18], as they govern a wide-ranging spectrum of applications of energetic XUV/x-ray sources. Multiple multi-XUV-photon-ionization by harmonic/attosecond sources provide an advanced tool for the study of ultrafast dynamics in correlated systems and/or of coupled motions due to the unprecedentedly short duration of their pulses, as well as for comprehensive temporal source diagnostics. Towards this goal a substantial increase of their pulse energy is required. Albeit FELs' markedly higher pulse energy (hundreds of μJ in the ~ 60 nm spectral range) [19], their capacity in ultrafast XUV science is determined by their so far longer pulse durations (tens of *fs*) [20] and shot-to-shot instabilities. Consequently, the enhancement of the HHG pulse energy in parallel with the reduction of the FEL pulse duration remain challenging tasks serving the same goal.

Here, using the high-harmonics generated by the interaction of high-power infrared (IR) *fs* laser pulses with Xenon (Xe) and Argon (Ar) gases, we demonstrate a 20-GWatt XUV source which delivers pulses with carrier wavelength $\lambda_{XUV} \approx 50$ nm and energy $E_{XUV} \approx 230$ μJ and $\approx 130$ μJ per pulse, respectively. Using this source, highly charged Ar ions (up to $Ar^{+4}$) have been observed (an accomplishment that was up to now feasible only with FEL sources [21]) and the process of multiple ionization of Ar has been investigated by measuring the dependence of the multiply charged ions on the intensity of the XUV pulses.

The 20-GWatt XUV source is based on the increase in the number of the XUV emitters using loose IR focusing geometry and the precise control of phase-matching conditions which was achieved by means of thin single-gas targets in a dual-jet configuration (supplementary material). The ≈ 18 m long beam line (Fig. 1a) was recently developed in the Attosecond Science and Technology (AST) laboratory of the Foundation of Research and Technology (FORTH). It is driven by a 10 Hz repetition rate Ti:sapphire laser system, which delivers $τ_L$ ≈ 20 fs pulses at 800 nm carrier wavelength (IR) and energy up to ≈ 400 mJ/pulse. A p-polarized IR pulse of 25 mJ - 45 mJ energy is focused by a spherical mirror of 9 m long focal length into the HHG area which hosts the non-linear medium provided by a dual-pulsed-jet (GJ1,2) configuration operated with the same noble gas (either Ar or Xe). The IR beam profile in the HHG area is shown in Fig. 1b. The generated harmonics reflected by a Silicon plate are spectrally filtered by means of either an Aluminum (Al) or a Tin (Sn) metal filter. The Sn filter was used for studying the multiple ionization of Ar, while the Al filter for measuring the pulse energy and the profile of the XUV beam. The harmonic spectra generated in Xe and Ar gases and the XUV beam profile after the Al filter are shown in Fig. 1c, 1d and 1e, respectively. For the studies of multiple ionization of Ar, the XUV radiation generated in Xe, after the Sn filter was focused by a gold coated spherical mirror of 5 cm focal length into an Ar gas-jet (Ar-GJ) placed in the target area. The harmonic spectrum at the position of the Ar-GJ is shown in Fig. 1f. The ionization products were measured by means of a magnetic bottle time-of-flight (MB-TOF) spectrometer that can be set to record either the photoelectron energy (PE) distribution or the ion-mass spectrum. In some cases (supplementary material) the harmonics passing through the filters have been, for convenience, also measured by recording the PE spectra resulting from the single-photon ionization of Ar after the interaction with the incoming (unfocused) harmonic beam. The energy of the XUV radiation in the generation region and the target area was determined by means of an XUV calibrated photodiode, taking into account the transmission of the filters and the reflectivity of the XUV optics.

The optimization of the generated energy in the dual-jet configuration was performed after maximizing the harmonic yield of the single-gas-jet (GJ1) (supplementary material). The energy maximizing conditions (GJ1 at the focus of the IR beam ($z = 0$), $P ∼ 25$ mbar, $L ∼ 1.5$ mm and $I_L$ just below the ionization saturation intensity) are in fair agreement with calculated values [22-24] (supplementary material Fig. S1). Single jet generation is limited by XUV absorption effects and IR-XUV phase mismatch induced by the neutral atoms and plasma generation in the medium which confines the coherent harmonic build-up to a short propagation length. This limitation can be overcome by applying quasi-phase matching conditions [22, 25] using two gas jets. The first one GJ1 is positioned at fixed $z \approx 0$ and the second one GJ2 at variable positions ($L$ and $P$ are the same in both jets). The dependence of the XUV energy generated by Xe and Ar gas on the distance between the two jets GJ2 and GJ1 is shown in Figs 2a and 3b, respectively. In both cases, the energy increases by a factor of $\approx 1.7$ when the GJ2 is placed at $z \approx \pm 5$ cm, verified by calculations taking into account the propagation effects in the dual-gas medium [22] (Fig. 2c). At this position, the generated XUV energy for Xe and Ar gas was $\approx 230$ µJ and $\approx 130$ µJ per pulse, respectively. The reduction of the energy around $z \approx 0$ is attributed to phase-mismatch effects induced due to the increase of the medium pressure and/or medium length, while the oscillations observed at $z > +5$ cm and $z < -5$ cm are due to the Gouy phase shift of the focused IR beam [26].

Taking into account the measured XUV pulse energy and a focal spot size of $\approx 2$ µm [27], when low ($\approx 90\%$) energy loss XUV optics are used, this radiation can support trains of *asec* pulses with overall duration $\tau_{XUV} = \tau_L/\sqrt{n} \approx 10$ *fs* (where $n$ = 3-5 is the order of non-linearity of the generation of plateau harmonic [28]) and $I_{XUV}$ up to $\sim 10^{17}$ W cm$^{-2}$. However, in the present work, for the investigation of multiple ionization process of Ar atoms the spectral region from 17.05 eV to 23.25 eV was selected (Fig. 1f) using XUV optical elements introducing $\approx 98\%$ losses. With the above given parameters, XUV intensities $I_{XUV}$ up to $\approx 7\times10^{15}$ W cm$^{-2}$ have been reached and multiply charged ions (Ar$^{n+}$) with $n$ = 1,2,3 and 4 have been observed (Fig. 3a). The dependence of the Ar$^{n+}$ (for $n \geq 2$) yield on $I_{XUV}$ is shown in Fig. 3b in log-log scale. In order to gain insight into the measured XUV intensity dependence of the Ar$^{2+}$ and Ar$^{3+}$ ion yields, we set up rate equations for considering ions up to Ar$^{4+}$. In the rate equations (supplementary material) we account for all energetically allowed processes, which are discussed below. For the two-photon-ionization of Ar we take the cross section equal to $10^{-51}$ cm$^4$s$^{-1}$, while for all other two-photon-ionization processes we use a cross section of $10^{-52}$ cm$^4$s$^{-1}$. However, we find that our results are robust for two-photon cross sections in the range from $10^{-51}$ cm$^4$s$^{-1}$ to $10^{-53}$ cm$^4$s$^{-1}$. For the three-photon processes we take the cross section to be equal to $10^{-85}$ cm$^6$s$^{-2}$. In Fig. 3b the points are the measured data, the dash-dot lines are fits in the experimental data and the solid lines are results of numerical calculations. The measured Ar$^{2+}$ signal after an increase with slope $s \approx 1.8 \pm 0.3$ saturates at $I_{SAT} \approx 2.2\times10^{15}$ Wcm$^{-2}$ where $s$ drops to $\approx 1.3 \pm 0.1$. The Ar$^{3+}$ signal has a slope $s \approx 2.9 \pm 0.5$ and saturates at $I_{SAT} \approx 3.2\times10^{15}$ W cm$^{-2}$ with $s$ dropping to $\approx 1.7 \pm 0.2$. The Ar$^{4+}$ signal appeared at intensities above the saturation intensity of Ar$^{3+}$ and was at the detection limit not allowing any intensity dependence measurement. The corresponding values from the numerical calculations are $s = 1.8$ and $I_{SAT} \approx 2.4 \times 10^{15}$ Wcm$^{-2}$ for Ar$^{2+}$ and $s = 3.1$ and $I_{SAT} \approx 3.7 \times 10^{15}$ Wcm$^{-2}$ for Ar$^{3+}$. Considering an uncertainty of a factor of $\approx 2$ (mainly associated with the measurement of the XUV focal spot size) in the estimation of the XUV intensity, the results are in fair agreement. Fig. 3c shows an excitation scheme that includes only the dominant processes.

The numerical calculations have access to and thus further provide significant information about the individual contributing ionization channels. In addition to the rate equations for the total ion yields, we compute the ion yields of each energetically allowed pathway contributing to the formation of Ar$^{2+}$ and Ar$^{3+}$. Fig. 4a shows these yields contributing to the Ar$^{2+}$ formation as a function of the XUV intensity. The black line is the yield of the two-photon direct double ionization (DDI) channel. All other curves denoted nPmP (m = 1, 2 and n = 2, 3) refer to the sequential pathways of *n*-photon ionization of Ar followed by m-photon ionization of Ar$^+$. Before saturation of the Ar$^+$ signal the dominant process for the Ar$^{2+}$ production is the two-photon DDI, while upon saturation the sequential processes set in and eventually prevails. In Fig. 4b several different channels producing Ar$^{3+}$ are compared. The notation hPiPjP (h=1,2, i=2, 3 and j=2, 3) stands for h-photon ionization of Ar, followed by i-photon DDI of Ar$^+$ or i-photon ionization of Ar$^+$ followed by j-photon ionization of Ar$^{2+}$. The dominating mechanism for the production of Ar$^{3+}$ is the single-photon ionization of Ar followed by a three-photon DDI of Ar$^+$. The slope of 4 of this four-photon process reduces to 3 in the intensity region of the experiment as in this region the Ar single photon ionization is saturated.

This work constitutes a demonstration that table-top HHG sources have reached intensities allowing the study and exploitation of highly non-linear XUV processes [29, 30], advancing the capacities in ultrafast XUV science [10, 27, 31]. The present results are highly relevant to the exciting prospect of the under implementation ELI-ALPS facility. The so called *GHHG SYLOS compact* beam-line [32] will use a similar arrangement having the additional advanced feature of operation at 1 kHz repetition rate. The XUV pulse energy levels of the present work at 1 kHz repetition rate and isolated pulse operation open up unique prospects for few body coincidence experiments in non-linear XUV laser-matter interactions exploited in detailed studies of ultrafast dynamics.

**Acknowledgments**

We acknowledge support of this work by the LASERLAB-EUROPE (grant agreement no. 284464, EC's Seventh Framework Programme), by the project "HELLAS-CH" (MIS 5002735) which is implemented under the "Action for Strengthening Research and Innovation Infrastructures", funded by the Operational Programme "Competitiveness, Entrepreneurship and Innovation" (NSRF 2014-2020) and co-financed by Greece and the European Union (European Regional Development Fund) and the European Union's Horizon 2020 research and innovation program under Marie Sklodowska-Curie grant agreement no. 641789 MEDEA. ELI-ALPS is supported by the European Union and co-financed by the European Regional Development Fund (GINOP-2.3.6-15-2015-00001). We thank G. Konstantinidis and G. Deligiorgis from the Materials and Devices Division of FORTH-IESL for their support in maintaining the quality of the optical components, N. Papadakis for developing the electronic devices used in the beam line, and S. Karsch from Max Plank Institute for Quantum Optics for his assistance on maintaining the performance of the Ti:S laser compressor. Also, A. Nayak (on leave from ELI-ALPS, presently employed at FORTH) and P. Tzallas (consultant member of ELI-ALPS), thank ELI-ALPS for the support and the fruitful collaboration.

**Figure Captions**

FIG. 1. (a) A drawing of the 20-GWatt XUV beam line. $SM_{IR}$: Spherical mirror of 9 m focal length. GJ1,2: Dual-pulsed-jet configuration placed on translation stages (TS). Si: Silicon plate. F: Al or Sn filter. $BP_{XUV}$: XUV beam profiler. $SM_{XUV}$: Gold coated spherical mirror of 5 cm focal length. Ar-GJ: Ar gas-jet. MB-TOF: Magnetic bottle time-of-flight spectrometer. $PD_{XUV}$: Calibrated XUV photodiode. FFS: Flat-field spectrometer. (b) IR beam profile at the focus. (c), (d) Harmonic spectra generated in Xe and Ar gases and transmitted by the Al filter. (e) XUV beam profile. (f) Harmonic spectrum generated in Xe gas transmitted by the Sn filter.

FIG. 2. Generation of 20-GWatt high-harmonics using a dual gas-jet. (a), (b) The upper panels shows the energy dependence of the XUV generated in Xe and Ar gas on the distance between GJ2 and GJ1 (placed at $z = 0$). The error bars represent one standard deviation. The lower panels show the harmonic spectrum measured in target area by recording the PE produced by the single-photon of Ar gas. (c) Calculated yield of the 17th harmonic generated in Ar as a function of the distance between the gas jets (black solid line). The yield was calculated for $I_L \approx 1.5 \times 10^{14}$ Wcm$^{-2}$ and $L \approx 1.5$ mm, $P \approx 25$ mbar for both jets. For comparison, the dependence of the XUV energy generated in a single Ar gas jet on the position of the GJ1 relative to the laser focus, is shown (black dashed-line).

FIG. 3. Multiple ionization of Argon atoms using the 20-GWatt high-harmonic source. **(a)** TOF mass spectrum produced by the interaction of the focused 11th-15th harmonics with Argon. The spectrum shows multiple charged Ar ions ($Ar^{n+}$) with $n$ up to 4. (b) Dependence of $Ar^{2,3+}$ yield on the $I_{XUV}$. For calibrating the XUV energy (x-axis) the $O_2^+$ signal was used. The black dashed lines show the linear fit on the raw data. The error bars represent one standard deviation of the mean. The solid lines show the results of the numerical calculations, including volume integration. As is expected from a single-photon ionization process, the dependence of the calculated $Ar^+$ yield on $I_{XUV}$ is linear and matches well with the experimental data of $N_2^+$. As the calculated $Ar^+$ yield (gray solid line) is orders of magnitude higher than $Ar^{2+}$, for visualization reasons, the calculated $Ar^+$ signal was divided by a factor of $8 \times 10^3$, and the measured $Ar^+$ signal was normalized to it. All, in arbitrary units measured, $Ar^{2+}$, $Ar^{3+}$ and $Ar^{4+}$ yield points are normalized to one and the same calculated single ion yield point. (c) Multi-XUV-photon multiple

ionization scheme (excluding higher order processes (ATI)) of Ar which supports the obtained results.

FIG. 4. Calculated ionization yields of different ionization pathways. These calculations are performed for a central photon energy of 22 eV, a pulse duration of 10 fs and no volume integration has been included. (a) shows the yield of $Ar^{2+}$ as a function of the XUV intensity. The black line is the yield of the two-photon DDI of Ar pathway. All other curves denoted nPmP (m=1,2, n =2,3) refer to the sequential pathways of n-photon ionization of Ar followed by m-photon ionization of $Ar^+$. From this graph we deduce that below saturation the two-photon DDI is the dominant pathway, while after saturation the lowest order sequential process prevails. (b) shows the XUV intensity dependence of the $Ar^{3+}$ yield. The black line is the single-photon ionization of Ar followed by three-photon direct ejection of two electrons from $Ar^+$, which is the dominant pathway at all intensities, while all other curves denoted hPiPjP refer to the sequential processes of h-photon (h=1, 2) ionization of Ar followed by i-photon ionization (i=2,3) of $Ar^+$ and eventually by j-photon ionization (j=2,3) of $Ar^{2+}$.

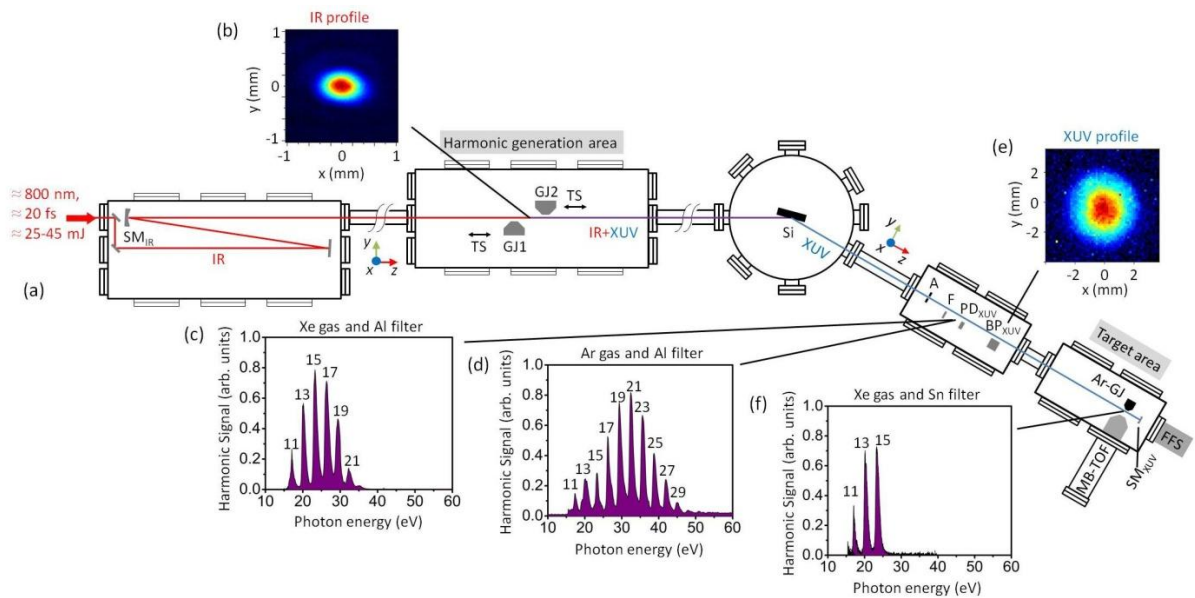

**Figure 1**

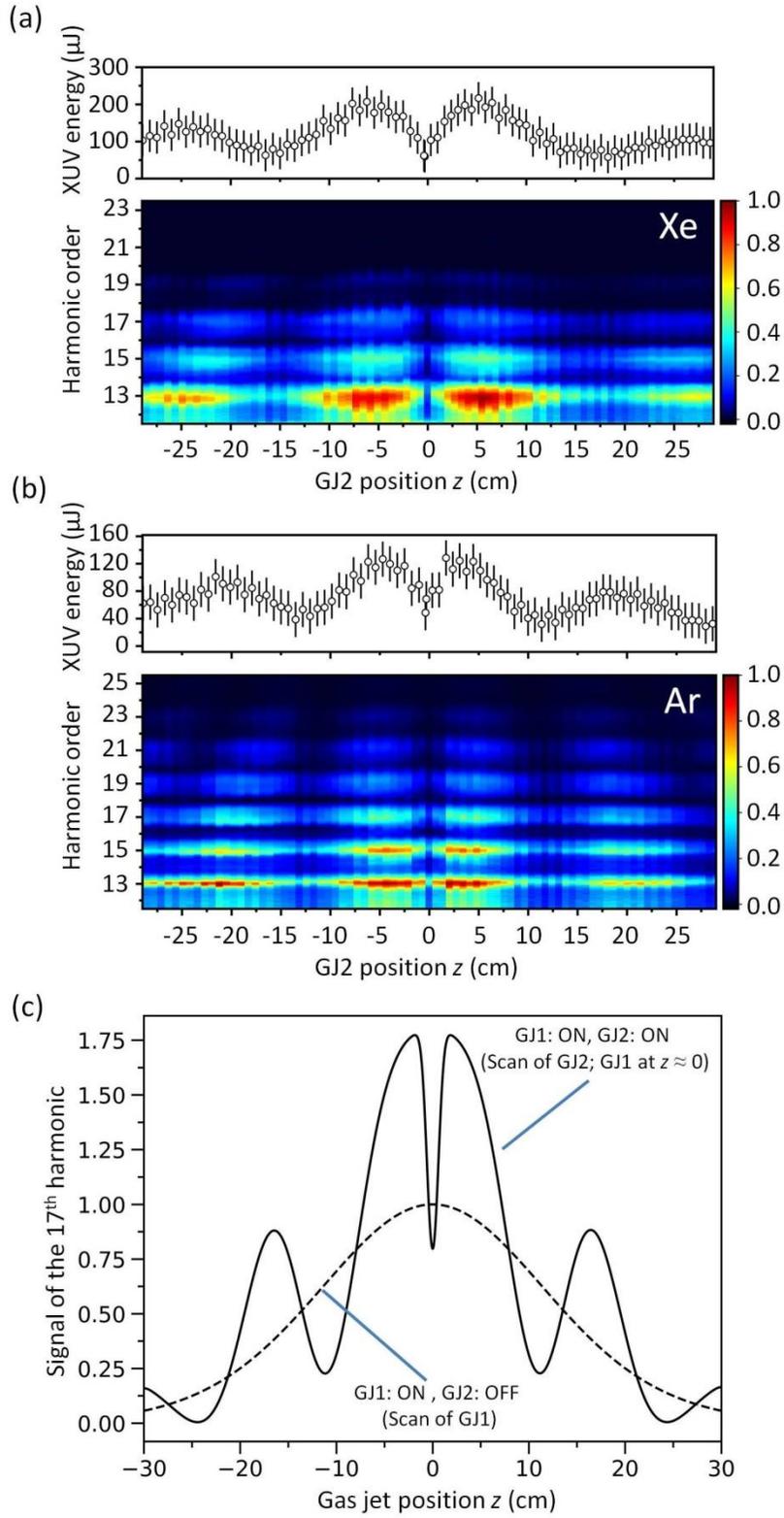

**Figure 2**

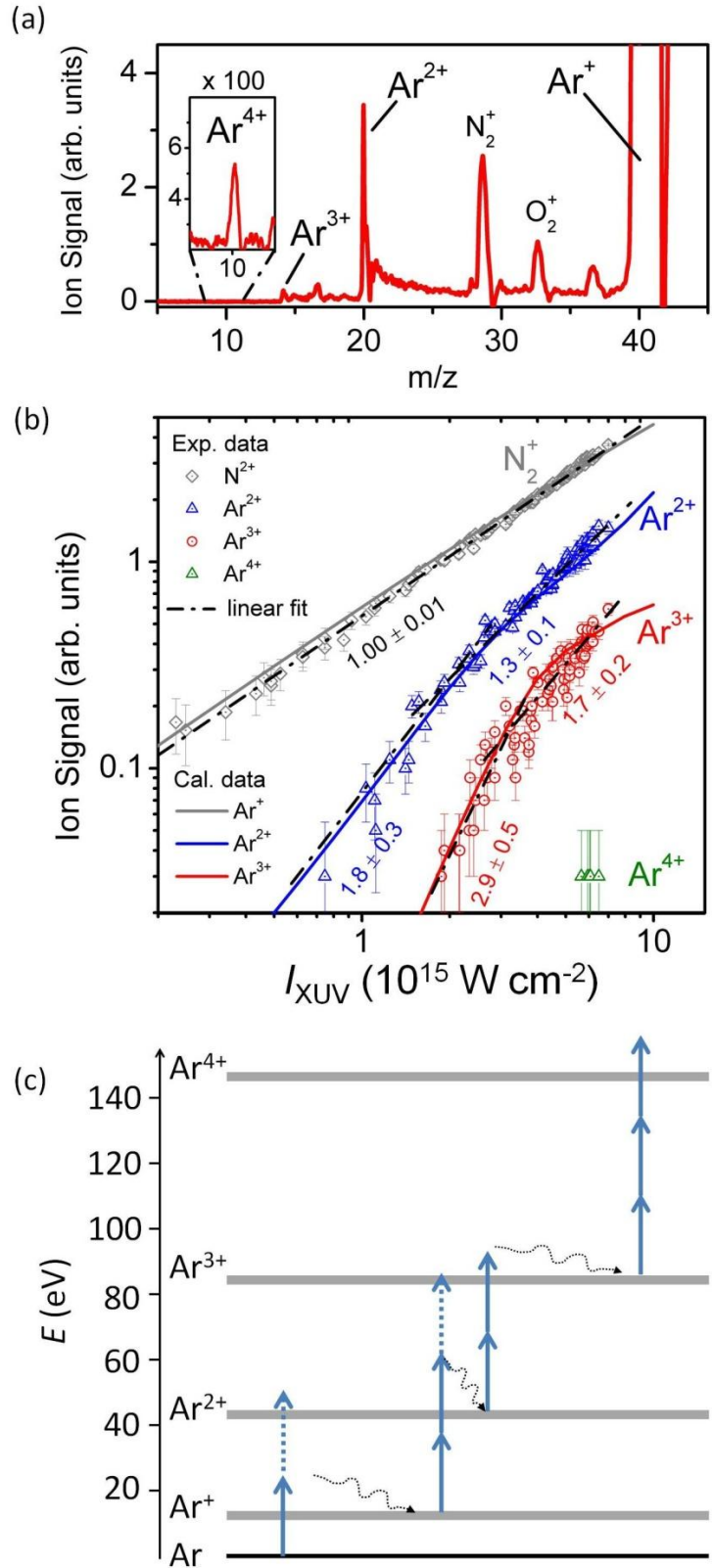

**Figure 3**

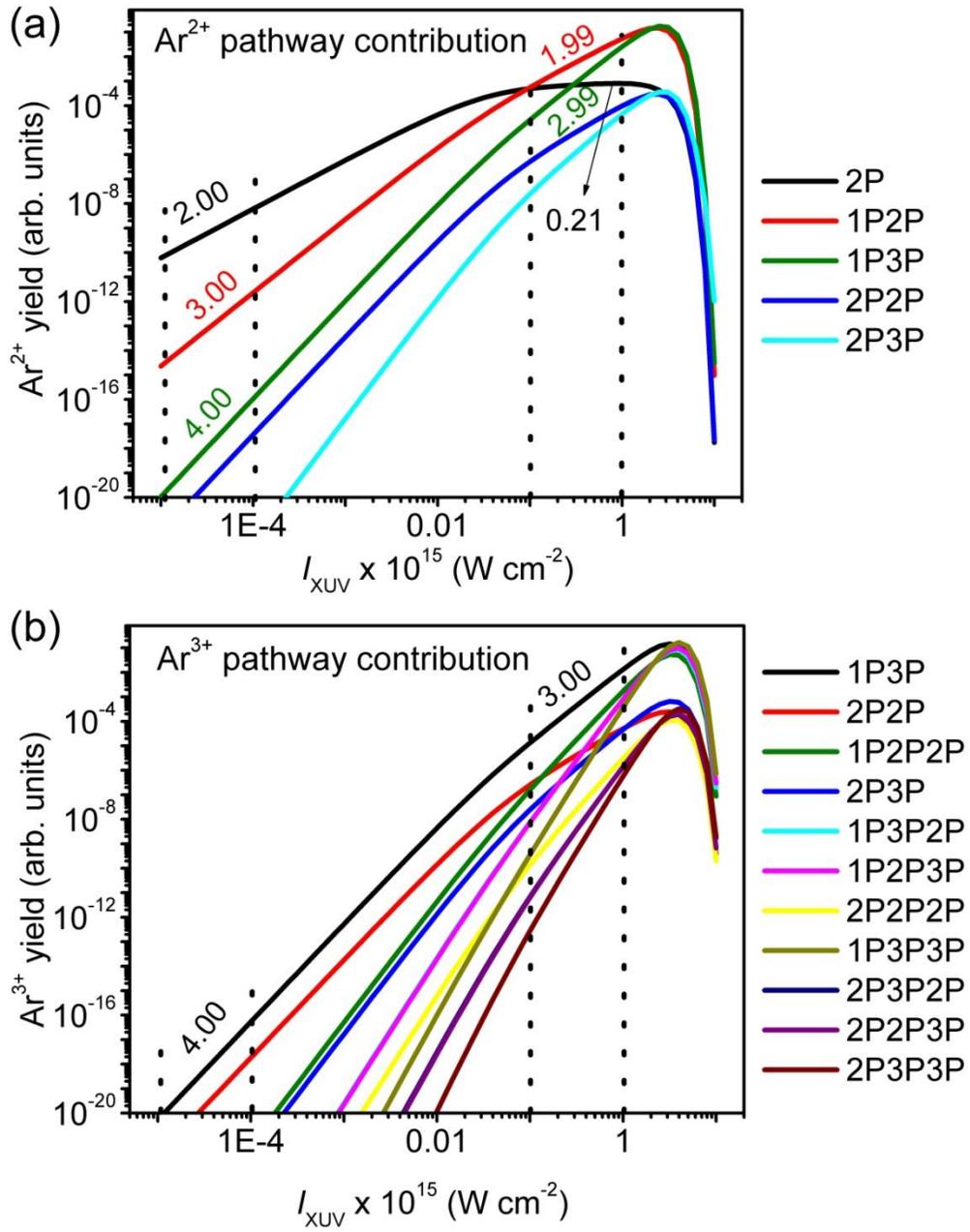

**Figure 4**

# Supplementary Material

**On the development of the 20-GWatt XUV source**

As mentioned in the main text, the operation principle of the 20-GWatt XUV source is based on the increased number of the XUV emitters and the precise control of phase-matching conditions, which was achieved by means of thin single-gas targets in a dual-pulsed-jet configuration with controllable distance between the jets. In gas-phase harmonics, the amount of the XUV energy exiting the gas medium is an interplay between the microscopic (single atom) and macroscopic (atomic ensemble) response of the medium. On the microscopic level, for a specific driving laser field wavelength ($\lambda_L$), the probability of the emission of a single XUV photon depends non-linearly on the driving laser field intensity ($I_L$) and the atomic properties. Considering that the probability of the single XUV photon emission is maximized for a fixed $I_L$ lying just below the ionization saturation threshold of the atom (which for Xenon and Argon atoms is $I_L < 3 \times 10^{14}$ W cm$^{-2}$), it is evident that for the enhancement of the energy of the XUV radiation one has to increase the number of the atomic XUV emitters and consider the macroscopic response of the medium. While keeping $I_L$, at the level of saturating single atom ionization the number of the emitters can be increased either by increasing the interaction volume (by increasing focal length together with laser pulse energy) or the atomic density of the medium. Incorporating the macroscopic response taking into account the propagation effects in the gas medium, it has been shown [ref. 22-24 of the main text and references there in] that for $L_{coh} \gg L_{abs}$ and $L_{coh} \gg L_{med}$ the XUV yield is proportional to $\propto (\rho \cdot L_{med})^2$. In the former expressions $L_{coh} = \pi/\Delta k$, $L_{abs} = 1/\rho\sigma^{(1)}$ and $L_{med}$ are the coherence length, the absorption length of the XUV radiation and the gas medium length, respectively, with $\Delta k = k_L - qk_L$, $q$ the harmonic order, $k_L$ the wave number of the driving field, $\rho$ the atomic density of the medium and $\sigma^{(1)}$ the single-XUV-photon ionization cross section of the atoms in the medium, and $A$ being the interaction area (or spot area of the driving field). This product constitutes the main scaling factor towards the enhancement of the produced energy.

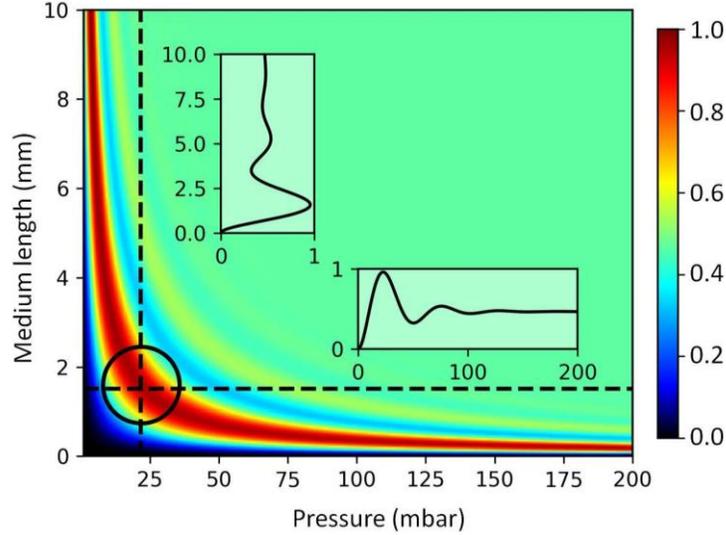

FIG. S1. Calculated harmonic yield generated in Ar gas as a function of the gas pressure ($P$) and medium length ($L$) for $I_L \approx 1.5 \times 10^{14}$ W cm$^{-2}$. The black circle depicts the conditions the values of $P$ and $L$ used in the experiment. The insets show a line-out of the harmonic yield along the dashed lines at $L \approx 1.5$ mm, $P \approx 25$ mbar.

Using one gas jet, for fixed $I_L$ the dependence of the harmonic yield on $P$ and $L_{med}$ is shown in the contour plot of Fig. S1. The red color area corresponds to the area of maximum XUV production and the black-circled area depicts the values of $P$ and $L_{med}$ used in the present work. The ~50% reduction of the XUV emission for "large"-length and "high"-pressure media (green color area in Fig. S1) is associated with the XUV absorption effects and IR-XUV phase mismatch induced by the neutral atoms and plasma generation in the medium which confines the coherent harmonic build-up to a short propagation length. This limitation can be overcome applying quasi-phase matching conditions as discussed in the main text.

In the experiment, the optimization of the generated energy in the dual-jet configuration was performed after maximizing the harmonic yield of the single-jet (GJ1). This was achieved by measuring the XUV energy as a function of the driving IR field intensity ($I_L$), the medium length ($L$), the gas pressure ($P$) and the position of the GJ1 relatively to the focus position of the IR beam. The optimum conditions were found when GJ1 was set to be at the focus of the IR beam ($z = 0$) where the IR was just below the ionization saturation intensity of Xe and Ar atoms, for $P \sim 25$ mbar and $L \sim 1.5$ mm (Fig. S2a, b). These optimal conditions are in fair agreement with the results obtained by the calculations (Fig. S1).

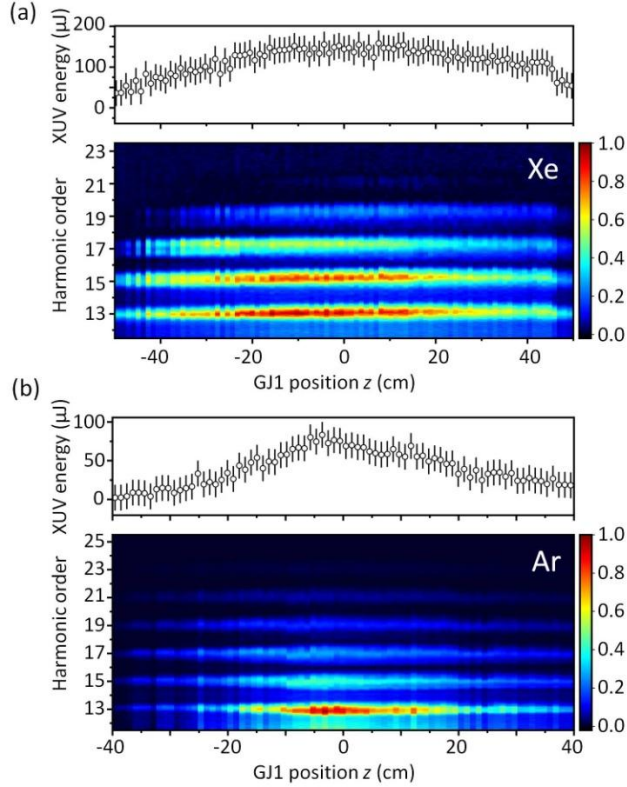

FIG. S2. (a) and (b) The upper panels show the dependence of the XUV energy (integrated over the spectrum passing through the Al filter) generated in Xe and Ar gas on the position of the GJ1 relative to the laser focus, respectively. For obtaining the energy values just after the harmonic generation area, the measured by the $PD_{XUV}$ energy values were divided by the reflectivity of the Si plate and the transmission of the Al filter. The error bars represent one standard deviation from the mean. The lower panels show the corresponding harmonic spectrum measured in target area by recording the single-photon PE spectra produced by the interaction of Ar gas with the incoming XUV beam.

In the dual gas-jet configuration, the gas-pressure and the medium length were the same for both gas-jets. This was confirmed by measuring the same harmonic yield generated by the individual jets when they were placed at the same position relative to focus of the IR beam. The piezo-based pulsed nozzle of slit shape orifice has dimensions 0.3 mm × 2 mm. The pressure and the medium length were estimated taking into account the backing pressure, the conductance of the orifice and its distance on the laser focus which was ~ 1 mm. The values are in agreement with those reported in ref. [1] where piezo-based pulsed nozzles have been used. The gas-jets were placed on x,y,z translation stages with the movement on the x and y-axes being manually controlled, while the displacement along the propagation z-axis was done by a motorized translation stage of ≈ 0.75 cm step (the minimum step of the stage was 5 μm). A Silicon (Si) plate, placed after the harmonic generation at the Brewster angle for the fundamental (i.e. 75deg), reflects the harmonics towards the detection area, while substantially attenuating the IR field. The reflectivity of the Si plate is ≈ 60% in the spectral range of 15 eV to 45 eV [2]. After reflection from the Si plate, the XUV radiation passes through a 5 mm diameter aperture (A) which blocks the residual outer part of the IR beam. The harmonics used in the experiments were selected by means of thin metal filters. An Sn filter with ≈ 20% transmission in the spectral range 17 eV - 23 eV was used for studying the multiple ionization of Ar, while for the measurement of the XUV energy and XUV beam profile a low transmission (≈ 5% in the spectral range 17 eV - 60 eV) Al filter was used in order to avoid damaging and/or saturating the detectors. The transmission of the filters was measured by recording the harmonic spectra with and without the filters. Also, the transmission of the IR beam after the metal filters was negligible. This was confirmed by the zero response of the $BP_{XUV}$ and $PD_{XUV}$ detectors (both sensitive to the IR radiation) after blocking the XUV beam by a 2 mm thick BK7 window. The harmonic spectra were measured by a flat-field spectrometer (FFS) attached to the back side of the target area chamber. The profile of the XUV beam was recorded by a detector ($BP_{XUV}$, consisting of a pair of MCPs and a phosphor screen) placed after the Al filter.

For the studies of the multiple ionization of Ar, the beam transmitted through the Sn filter was then focused by a gold coated spherical mirror ($SM_{XUV}$ positioned at ~2° angle of incidence) of 5 cm focal length into the target area where the Ar-GJ is placed. The reflectivity of the gold mirror (≈ 12%) is constant in the spectral region of 17 eV - 23 eV [3]. We note, that in the spectral range from 15 eV to 30 eV, the measured photoelectron distribution does not differ significantly from the spectrum measured by the FFS as in this photon energy range the single-photon-ionization cross section of Argon is almost constant at ≈ 30 Mb [4]. The deviation (compared to the spectra recorded by the flat-field spectrometer) appearing at photon energies > 30 eV is attributed to the reduction of the single-photon ionization cross-section.

In order to estimate the conversion efficiency ($C_{(q)} = E_{XUV}^{(q)}/E_{IR}$) of the harmonic generation process the energy per harmonic ($q$) per pulse ($E_{XUV}^{(q)}$) was calculated taking into account the transmission of the filter, the reflectivity of the Si plate and the quantum efficiency of $PD_{XUV}$. For the single-jet configuration, and for the optimum generating conditions, it has been found that the maximum generated XUV energy (integrated over the spectrum) was ≈ 135 µJ and ≈ 75 µJ per pulse for Xe and Ar, respectively. This corresponds to an $E_{XUV}^{(q)}$ ($q$=11, 13, 15) ≈ 30 µJ and ≈ 10 µJ for Xe and Ar gas, respectively, considering that the total energy is shared equally between the plateau harmonics. This results to $C_{(11)} \approx C_{(13)} \approx C_{(15)} \approx 1 \times 10^{-3}$ and ≈ 2 × $10^{-4}$, respectively. The enhancement by factor of > 2 as compared to previously reported values [5] (where $C_{(11)} \approx 5 \times 10^{-4}$) is associated with the increased interaction volume and is in agreement with the energy scaling law reported in ref. 6 and ref. 24 of the main text. The dual-jet configuration provides a factor of ≈ 1.7 further enhancement, resulting to a conversion efficiency ≈ 2 × $10^{-3}$ and ≈ 3 × $10^{-4}$, for Xe and Ar, respectively.

The signal of the single-charged ions ($Ar^+$, $O_2^+$, $N_2^+$ etc) shown in Fig. 3a of the main text of the manuscript, is proportional to the XUV pulse energy as it is generated by a single-photon ionization process. These ions are mainly produced away from the focus into the target area. The singly-charged ion signal produced at the focus of the beam is considered negligible compared to the signal of the singly charged ions produced outside of the focus due to volume and single-photon ionization saturation effects which according to the Lowest-Order-Perturbation theory are taking place for $I_{XUV} > I_{XUV}^{(sat)} \approx 7 \times 10^{12}$ W cm$^{-2}$) [3].

**On the numerical calculations for obtaining the multiple charge Argon ion yield**

In the numerical calculations in obtaining the ion yields and the contribution of different pathways to the final ions, we use a Gaussian laser pulse of FWHM equal to $\tau_{XUV}$ = 10 fs. Moreover, we perform a volume average for our results for the ion yields. We do so as follows. First, for a certain peak intensity we used the following equation [7] to determine the intensity, $I_{XUV}$, at each point ($r$, $z$) in cylindrical coordinates:

$$I_{XUV}(r,z;t) = I_{XUV}(t) \frac{w_0^2}{w(z)^2} exp\left[-\frac{2r^2}{w(z)^2}\right] \quad (1)$$

where $r$ is the radius and $z$ is the beam propagation axis. $w(z)$ is the beam radius, defined in terms of the beam waist, $w_0$ = 1 µm, and the Rayleigh length, $z_R$ = 51.5µm as,

$$w(z) = w_0\sqrt{1 + (z/z_R)^2} \quad (2)$$

We calculated the ion yield in a volume with limits $r_{max}$ = 3 mm in the radial direction and $z_{min}$ = -3 mm to $z_{max}$ = 3 mm in the $z$ direction. These ion yields were then integrated using the following expression [8]:

$$P_i = \int_0^{r_{max}} \int_{z_{min}}^{z_{max}} 2\pi r N_i(r,z) dz dr \qquad (3)$$

where $P_i$ is the yield of the ion $i$ integrated over the volume, $N_i$ is the yield of ion $i$. We have checked that our results for $P_i$ converge. The yields for the main pathways leading to the formation of $Ar^{2+}$ and $Ar^{3+}$ are not volume integrated. In our computations the 3 photon transition is energetically allowed when the photon energy is equal to or above 22 eV. We find that our results do not change when the photon energy changes from 22 eV to 23.3 which is the maximum energy considered in the experiment. However, if we reduce the photon energy below 22 eV then the 1P3P process is not energetically allowed and the slope we obtain for the $Ar^{3+}$ ion yield is closer to 4. We note that in our computations all energies are obtained with MOLPRO (a quantum chemistry package) using Hartree-Fock with a 6-311G basis. We have used the same basis for previous FEL processes in Ar [9].

**Methods References**